\documentstyle[twoside,fleqn,espcrc2]{article}

\newcommand{\AmS}{{\protect\the\textfont2
  A\kern-.1667em\lower.5ex\hbox{M}\kern-.125emS}}

\newcommand{\beq}{\begin{equation}}
\newcommand{\eeq}{\end{equation}}
\newcommand{\bea}{\begin{eqnarray}}
\newcommand{\eea}{\end{eqnarray}}

\newcommand{\nonu }{\nonumber}

\newcommand{\Ds}{\not\!\! D}

\newcommand{\aas}{/\kern-.50em A}

\newcommand{\ra}{\rightarrow}

\title{Instanton effects and chiral symmetry breaking in QCD2}

\author{Hugo R. Christiansen\thanks{E-address: hugo@cat.cbpf.br}
\address{Centro Brasileiro de Pesquisas 
F\'\i sicas,  CBPF, \
Department of Fields and Particle Physics, DCP\\
Rua Dr. Xavier Sigaud 150, 22290-180\, Rio de Janeiro, Brazil}}

\begin{document}

\begin{abstract}
In this paper we discuss the spontaneous breakdown of chiral symmetry  
in Quantum Chromodynamics by considering gluonic instanton configurations
in the partition function. 
It is shown that in order to obtain nontrivial fermionic 
correlators in a two dimensional gauge theory for the strong interactions
among  quarks, a regular instanton background has to be taken into account. 
We work over massless  quarks in the fundamental representation 
of SU$(N_c)$. For large $N_c$,  massive quarks are also considered.
\end{abstract}

\maketitle

\section{INTRODUCTION}
In the massless limit, 
Quantum Chromodynamics is not only invariant under right (${}_R$) and 
left (${}_L$) chiral symmetry transformations 
but also under two independent 
Abelian symmetry groups U$(1)$. The massless theory is thus symmetric 
under a large unitary group U$(N_f)\times$U$(N_f)$. The vector subgroup 
is realized in the normal mode, i.e. the QCD vacuum is also invariant 
under U$_{L+R}(N_f)$ transformations. The hadrons form degenerate 
SU$(N_f)$ multiplets for small $N_f$ and baryon number is conserved. 
The remaining symmetries -axial SU$(N_f)$ and axial U(1)- corresponding to 
U$_{L-R}(N_f)$, are not manifest in particle degeneracies as we do not 
observe parity doubling of baryon states. This phenomenom is ascribed 
to the
vacuum of the theory which is said to spontaneously breakdown the axial
symmetry (SSB). In a three flavor world it has been experimentally verified
that SSB is true for the special part of the axial group. Actually, eight 
approximately massles pseudoscalar mesons are observed in the hadronic 
spectra.
However, there is one missing particle which would be associated with 
U$_A(1)$ according to the Goldstone theorem, since the  
transformation $\psi_f \rightarrow e^{i\gamma_5\theta} \psi_f$ is
neither realized as a symmetry of the vacuum nor \`a la Goldstone.
Although we do not exactly know the essential physical mechanism 
underlying chiral symmetry breaking,
the answer to this famous problem lays on the so-called
Adler-Bell-Jackiw anomaly.
This explicit symmetry breaking at the quantum level 
has a close connection to 
a, seemingly,  topological nature of the QCD ground state.

In two dimensions, it turns out that there is no possible Goldstone 
mode associated with any continuous (global) symmetry of the Lagrangian 
\cite{col}. If chiral symmetry cannot be broken by 
the vacuum state and parity is a good symmetry of the Lagrangian, 
the 2D spectrum should contain parity doublets. However, they do not appear,
in direct correspondence with the 4D situation. The mechanism which 
eliminates the pairing among meson states will be clarified in a 
path-integral calculation. We will see that a non-Abelian chiral
anomaly takes place provided a topological structure is taken into account. 
In this way, not only the Abelian part of the axial symmetry but also the 
non-Abelian counterpart are dynamically broken due to quantum fluctuations. 
It shall be made manifest by performing a fermion chiral decoupling. 

The non-perturbative vacuum structure of QCD can be expressed in terms of 
nonzero v.e.v. of various composite operators. These condensates have 
been introduced as phenomenological parameters in a non-perturbative 
generalization of the operator product expansion, which can be related 
to observable hadronic properties by the sum rule method. The lowest 
dimensional condensates are well-known phenomenologically, but 
considerable uncertainty prevails about 
the value of higher dimensional correlators. Condensates can be understood 
as being generated by certain nonperturbative fluctuations of the fields. 
In particular, instantons would be responsible for the SSB of chiral 
symmetry.
A quark condensate arise due to the delocalization of fermion zero 
modes associated with the instantons in the medium 
\cite{dia-pe}.

Therefore, the study of fermionic correlation functions is a key issue
of Quantum Chromodynamics as these are suitable quantities to shed 
light on QCD non-perturbative aspects and hadron physics. Actually,
the correlation among quark fields can be written in terms
of dispersion relations involving matrix elements among the vacuum and
physical hadronic Fock states. Consequently, fermionic correlators
of fundamental degrees of freedom are directly connected to the normalized 
cross sections of hadronic reactions \cite{pheno}. 
In this way, using hadronic phenomenological data,
valuable information about the fundamental structure of matter
can be obtained from the anlysis of multipoint fermionic correlators.  
The QCD vacuum has non-perturbative condensation of quarks and gluons. 
This fact is extracted from the studies of current algebra, sum rules and 
lattice QCD.
For instance, a non-perturbative structure of the QCD vacuum comes about 
from the Gell~Mann-Oakes-Renner relation, 
\[
\langle \bar u u+\bar d d\rangle =-{2f_{\pi}^2 m_{\pi}^2}/{(m_u+m_d)},
\]
since such a nonzero result indicates the existence of a 
dynamical mass in the massless quark propagator which would  vanish
in the perturbative calculation.
The strong attractive force in the $J^P=0^+$ channel induces the 
instability of the Fock vacuum of massless quarks to realize the 
nonperturbative ground state with quark antiquark condensation. 
This is quite similar to the BCS mechanism of superconductivity. 
The pair condensation breaks the full symmetry down to the vector 
subgroup and the dynamical quark mass is generated.

Therefore, the study of fermionic condensates is especially useful 
in connection with chiral symmetry breaking coming from non-perturbative 
effects, seemingly arising due to an underlying topological structure.

\section{MODEL}

These highly complicated problems can be faced only by assuming sensible 
simplifications in the theory itself, or for instance, on the number of 
space-time dimensions.
Quantum chromodynamics in two dimensions is a convenient framework
for several reasons: the use of fundamental degrees of 
freedom, its non-Abelian character, the chirality properties of the theory, 
the existence of analytical 
results, etc. We will then work with the following Lagrangian 
of SU$(N_c)$ gauge fields coupled to Dirac fermions in an Euclidean space:
\[
L=\bar\psi^{q}(i\partial_{\mu} \gamma_{\mu} \delta^{qq'}+A_{\mu,a}
 t_a^{qq'}\gamma_{\mu})\psi^{q'}+
\frac{1}{4g^2} F_{\mu\nu}^a F_{\mu\nu}^a 
\]
where $q$ runs from 1 to $N_c$ and $a$, the label of the generators, 
goes from 1 to $N_c^2-1$. For short, we will 
discuss in detail the issues for only 
one flavor, but the procedure can be readily extended to arbitrary $N_f$.
From now on we drop the ${}_c$ subindex from $N_c$.

In order to compute fermionic correlators, a path-integral approach 
is very appropriate, especially to handle different topological sectors. 
A necessary
first point in the procedure is to separate the gauge field as a sum of a 
fixed classical background carrying a topological charge $n$, and a quantum 
fluctuation belonging to the trivial sector: 
$A_\mu^a(x) = A_\mu^a{}^{(n)} + a_\mu^a$. 
In this way one is able to decouple gauge fields from fermionic fields within 
the $n$=0 sector, yielding a non-Abelian jacobian 
of the Fujikawa type \cite{fuji}.
In order to do this in a gauge independent way we have to use a group valued
representation for all the fields and then perform a chiral rotation of the 
spinors.
It is worthwhile to show the  jacobian resulting from the
decoupling, since it first makes apparent the axial anomaly,
responsible for the full chiral symmetry breaking in 2D. 
In terms of group fields $u,v$ and $d$ which represent $a_+, a_-$ and 
$A^{(n)}_+$ respectively ($A^{(n)}_-=0)$,  
the fermion determinat can be suitably 
factorized, in an arbitrary gauge,
by repeated use of the Polyakov-Wiegmann identity,
resulting in the following formal identity for the jacobian
\[
J=\frac{\det \Ds[A^{(n)} + a]} 
 {\det \Ds[A^{(n)}]} ={\cal N} \exp{-S_{eff}(a, A^{(n)})}
\]
where
\begin{eqnarray}
&& S_{eff}(a, A^{(n)})  =  W[u] + W[v] \nonu\\
&& +\frac{1}{4\pi}tr_c\int d^2x (u^{-1} 
\partial_+ u) (d \partial_- d^{-1}) \nonu \\
&&  +\frac{1}{4\pi}tr_c\!\int\! d^2x\, (d^{-1} \partial_+ d) 
(v \partial_- v^{-1})\nonu\\
&& +\frac{1}{4\pi}tr_c\!\int\! d^2x\, 
(u^{-1} \partial_+ u)\, d\, (v \partial_- v^{-1})\, d^{-1},
\nonumber
\end{eqnarray}
$W[u]$ being the usual Wess-Zumino--Witten action.
A lengthy calculation leads to the exact expressions for the
correlation functions of an arbitrary number of fermionic bilinears 
in the SU$(N)$ 2D gauge theory \cite{ijmp}. As it is a very large expression
we shall not quote it in this note but we can state here that our 
results show that the simple product of fermionic and bosonic path-integrals
one finds in the Abelian case \cite{hf1} becomes here 
an involved sum due to color couplings, bringing about the extended 
Wess-Zumino-Witten action coming from the decoupling.
Nevertheless, we can say that expressions show in a 
clear way both sectors completely decoupled.
The fermionic path-integral can be easily performed, amounting
to a sum of products of zero modes of the Dirac operator in the 
multinstanton background. 
In the bosonic sector, the presence of the 
Maxwell term crucially changes the effective dynamics with respect
to that of a pure Wess-Zumino model. One then has to perform 
approximate calculations  to compute the bosonic factor,
for example, by linearizing the group transformations;
nevertheless, the point relevant to our discussion of
obtaining an nonzero ferm\-ionic correlators is manifest in our result. 

We should note however that an immediate by-product of our approach 
gives a null elementary (lowest dimensional) condensate, 
$\langle \bar\psi\psi\rangle = 0$, for any number of colors.
For finite $N$, it is consistent with alternative approaches, but,
on the other hand, independent analytical calculations
coming from dispersion relations and canonical quantization \cite{other} 
have found that it is possible to obtain a nonzero condensate 
when the large $N$ limit is considered in massive QCD2. 
These alternative approaches, which consider just trivial topology, 
would in principle be in constrast with Coleman's theorem as we 
have discussed in the
introduction. However, it has been shown that in the weak
coupling regime, or 't Hooft phase, there is in fact a 
Berezinskii-Kosterlitz-Thoules
(BKT) \cite{bkt} realization of the chiral symmetry, defining a 
phase transition  between a chiral symmetric and a chirally broken
phase. 
What we want to point out now is that by considering 
the massive version of our model for large $N$, 
we may also obtain a nonzero outcome for $\langle\bar\psi\psi\rangle$ in
the chiral limit.

Regarding the topological aspects of our approach, 
here we are considering fundamental quarks in a background 
given by the gluonic component of 
Z$_N$ vortices \cite{sha-dv} which give a realization in an 
Euclidean 2D theory of regular instanton configurations.
This picture allows an interesting connection between an underlying 
topological structure in the theory and the existence of
nonzero fermion condensates. 
This is a very suitable feature for a model approach to the
strong interactions among quarks. The relevance of instantons has 
been amply demonstrated
by phenomenology as well as by lattice calculations. In particular,
instantons allow a good description of chiral symmetry breaking 
and shed light on the nonperturbative
phenomena determining the structure of hadrons. An instanton vacuum 
certainly provides a convenient tool for computing correlation functions
and a microscopic picture of the nonperturbative 
configurations of the gluon field.

The minimal correlation function, 
for the  massive partition function, can be readily written as

\bea
& & \langle\bar\psi\psi(\omega)\rangle_M =\sum_n\bigg(\
\langle\bar\psi\psi(\omega)\rangle_{M=0}^{(n)} + \\ 
& & 
M \int d^2x \langle\bar\psi\psi(\omega)\bar\psi\psi(x)
\rangle_{M=0}^{(n)} + \frac{1}{2}\, M^2 \nonumber\\
& & \left.\int d^2x\, d^2y
\, \langle\bar\psi\psi(\omega)\bar\psi\psi(x)
\bar\psi\psi(y)\rangle_{M=0}^{(n)} + \dots \right )\nonumber 
\label{cm2}
\eea
Written in this fashion, it is apparent that for $M\neq 0$ the elementary 
condensate receives contributions from every higher order
correlator coming from the massless
theory. As we have mentioned, these have precisely been calculated 
in a general way \cite{ijmp}. It follows from the discussion above 
that these correlators are given by the fermion zero
modes determined by the topological background.

Since the fermionic sector is completely decoupled, the counting 
of the nonzero terms simply follows from that of the Abelian case 
because the topological structure here can be red out from the 
torus of the gauge group. To be more specific, 
in a compactified space, there exist  $n N$ normalizable zero 
modes in topological sector $n$ \cite{we}; this implies
the vanishing of the first summatory in eq.(1) 
in every topological sector. 
However, for higher powers of $M$ it is clear that certain nonzero
contributions come into play. 
The zero modes, with
a definite chirality, set the integration rules for computing v.e.v's.
By using the chiral decomposition 
$\langle \bar\psi\psi(x)\rangle =\langle \bar\psi_R\psi_R(x)
\rangle+ \langle\bar\psi_L\psi_L(x) \rangle$
one can see that the first $N$ powers of $M$
($j= 1\dots N$) receive an input from the trivial 
topological sector alone.
For $j\geq N$, the $n=1$ sector starts contributing 
together with $n=0$.
For higher powers, $j\geq 2N$,  the contribution of topological 
sector $n=2$ starts on, etc.
Now, it can be easily seen that the number of contributions grows 
together with the number of colors. 
As we let $N$ go to infinity the elementary  condensate in the massive
theory does so.
On the other hand, since within each nontrivial topological sector
the number of zero modes grows also to infinity, one has
divergent v.e.v. everywhere in the series expansion of eq.(1). 
Accordingly, high order terms can also produce a nonzero outcome
in the chiral limit.
This is a pleasant result of our model, in order to mimic real 
four dimensional QCD where an instanton vacuum seems to be closely
connected to  this  phenomenum.

Therefore, it is clear that the limit $M\ra 0$ becomes matter of a
careful analysis; namely, combined with a large number
of colors,  eq.(1)  leaves place 
enough for a nontrivial elementary condensate even in the chiral limit. 
This result is of 
phenomenological interest as it puts forward possible roots to the BKT
phenomenom,  arising from topological considerations in a 
path-integral approach. 

\section{CONCLUSION}
We have discussed chiral symmetry breaking in a
two dimensional model for the gluon interactions among 
quarks. Our approach allows a systematic procedure
for computing arbitrary  correlation functions of fermionic operators.
We have shown that a topological background is crucial for obtaining 
a whole class of nonzero condensates of the fundamental fields
which signal the dynamical breakdown of the full chiral symmetry group.

\subsection*{Acknowledgements} 
I would like to thank F.A.~Schaposnik 
for enlightening discussions.  This work was supported by DCP-CBPF
and FAPERJ, Brazil. Thanks are also due to S.~Narison, 
chairman of the Conference.

\end{document}